# Introducing the anatomy of disciplinary discernment: an example from astronomy

Urban Eriksson[1,2], Cedric Linder[1,3], John Airey[1,4], Andreas Redfors[2]
[1]*Department of Physics and Astronomy, Uppsala University, Uppsala, Sweden*
[2]*School of Education and Environment, Kristianstad University, Kristianstad, Sweden*
[3]*Department of Physics, University of Western Cape, Cape Town, South Africa*
[4]*Department of Languages, Linnaeus University, Kalmar, Sweden*
For correspondence: urban.eriksson@hkr.se

**Abstract :**
Education is increasingly being framed by a competence mindset; the value of knowledge lies much more in competence performativity and innovation than in simply knowing. Reaching such competency in areas such as astronomy and physics has long been known to be challenging. The movement from everyday conceptions of the world around us to a disciplinary interpretation is fraught with pitfalls and problems. Thus, what underpins the characteristics of the disciplinary trajectory to competence becomes an important educational consideration. In this article we report on a study involving what students and lecturers discern from the same disciplinary semiotic resource. We use this to propose an Anatomy of Disciplinary Discernment (ADD), a hierarchy of *what* is focused on and *how* it is interpreted in an appropriate, disciplinary manner, as an overarching fundamental aspect of disciplinary learning. Students and lecturers in astronomy and physics were asked to describe what they could discern from a video simulation of travel through our Galaxy and beyond. In all, 137 people from nine countries participated. The descriptions were analysed using a hermeneutic interpretive study approach. The analysis resulted in the formulation of five qualitatively different categories of discernment; the ADD, reflecting a view of participants' competence levels. The ADD reveals four increasing levels of disciplinary discernment: Identification, Explanation, Appreciation, and Evaluation. This facilitates the identification of a clear relationship between educational level and the level of disciplinary discernment. The analytical outcomes of the study suggest how teachers of science, after using the ADD to assess the students disciplinary knowledge, may attain new insights into how to create more effective learning environments by explicitly crafting their teaching to support the crossing of boundaries in the ADD model.

> *I could not help laughing at the ease with which he [Sherlock Holmes] explained his process of deduction. "When I hear you give your reasons," I remarked, "the thing always appears to me to be so ridiculously simple that I could easily do it myself, though at each successive instance of your reasoning I am baffled until you explain your process. And yet I believe that my eyes are as good as yours." "Quite so," he answered, lighting a cigarette, and throwing himself down into an armchair. "You see, but you do not observe"*
>
> *Exchange between Sherlock Holmes and Dr. Watson*
> *A Scandal in Bohemia*
> *Arthur Conan Doyle (1891)*

**Introduction**

At any one point in time a myriad of information is available to us through our senses, however we can only ever focus on a small portion of this information at one time (Medina, 2008). In the dialogue above, the master detective, Sherlock Holmes, not only focused on different things than Dr Watson, he also interpreted the available information differently. Put simply, from the multitude of sensory input available to him, Sherlock Holmes had learned what was important to focus on for the case at hand and how to assign meaning to it. In this article, what Sherlock Holmes refers to as observation we are calling discernment (Eriksson, Linder, Airey, & Redfors, 2014). We characterize discernment in terms of coming to know what to focus on and how to appropriately interpret it for a given context. Since the value of knowledge today lies much more in competence performativity and innovation (see, for example, Gilbert, 2005), we make the case that becoming competent in any discipline involves a similar process, namely learning; *what* to focus on in a given situation and *how* to interpret it in an appropriate, disciplinary manner. We use data drawn from university astronomy to generate an anatomy of disciplinary discernment (ADD)—a hierarchy of what is focused on and how it is interpreted. We go on to conclude that (1), there is a mismatch between what lectures discern and what students discern, (2), the ADD can be used to assess student competence development, (3), the role of the teacher in disciplinary learning may be framed in terms of helping students to cross category boundaries in the ADD.

For students who are entering a discipline, the appropriate disciplinary interpretation of any given input is often as impenetrable to them as Sherlock Holmes' discernment was to Dr Watson. Learning astronomy has been shown to be particularly challenging for many students (see, for example, Bailey, Prather, Johnson, & Slater, 2009; Sadler, 1996; Vosniadou & Brewer, 1994; Wallace, Prather, & Duncan, 2012). Why is this so? We propose that, like Dr Watson, aspirant astronomers need to learn how to sort through the myriad of information that any aspect of astronomy presents in a disciplinary manner. Thus, we chose to use astronomy as the educational environment in which to situate our study. The study involved asking students and lecturers to describe what they discern from an information-rich simulation video of a journey through our Galaxy and beyond (Tully, 2012). These descriptions were used to formulate an Anatomy of Disciplinary Discernment (ADD) –the ways in which the disciplinary meanings of a given representation may be discerned.

**Disciplinary discernment – an overview**

Disciplinary discernment is a key concept for this article and so we present a summary of pertinent work done to date on characterizing learning in terms of *noticing* and *reflection* — concepts that we use to define disciplinary discernment.

We begin with noticing. When learning something new one needs to either re-work existing knowledge or make some kind of observations that trigger new ideas. These observations, or experiences come about through perception, in our case through visual perception; by the *noticing* of something. Therefore, noticing is an important first step in any learning process (Mason, 2002). Noticing is something that humans do all the time. It is an unconscious act that we cannot deliberately choose to do or not to do (Mason, 2002). Our senses provide information to our brain that we process, usually in an unconscious way, and only some of this information comes to our conscious awareness. This is what we notice. To notice something is thus to distinguish it from the background or surroundings, leading to changes in our perceptual system, our brain. Hence, 'we notice all the time, but on different levels' (Mason, 2002, p. 33). However, most of what is noticed is lost from accessible memory. To remember what is noticed calls for a *marking* in the memory of that noticing. One can then come back to this marking and re-mark it for future use, often in combination with reflection (Schön, 1983, see below). The next level of the noticing process would be the *recording* of the noticing. Here one makes, for example, a written note that one can come back to and through more reflection, putting the perceived noticing into focal awareness in order to construct meaning (Marton & Booth, 1997). Such meaning-making is likely to change one's thinking (Mason, 2002, p. 34). For us, this construction of meaning characterises the process of learning in terms of discernment. However, the discernment is different for different persons depending on their background, past experiences and disciplinary educational level (Latour & Woolgar, 1979;Latour 1986). Lindgren and Schwartz (2009) refer to this as *the noticing effect* and define it as follows:

> 'A characteristic of perceptual learning is the increasing ability to perceive more in a given situation. Experts can notice important subtleties that novices simply do not see…[This] helps explain how people can come to perceive what they previously could not, and how the ability to notice often corresponds to competence in a domain.' (p.421)

Marton and Booth (1997) have argued that it is not possible to simply get oneself to notice relevant 'things'. They have proposed what is known as the variation theory of learning, which has been shown to be useful as a way to overcome this difficulty (see, for example, Bernhard, Carstensen, & Holmberg, 2007; Fraser & Linder, 2009). Humans need exposure to appropriate variability to promote the noticing. Variability is thus 'essential for learning to notice what is important *and* what is not important' (Lindgren & Schwartz, 2009, p. 426, emphases in the original), leading to what Goodwin (1994) has called *professional vision*. Appropriate experiences thus enable humans to notice more in a given situation.

We now turn to reflection. Noticing is a natural and spontaneous process for all humans (Mason, 2002). However, what both Mason (2002) and Lindgren and Schwartz (2009) emphasize is that the concept of noticing is not sufficient in itself to explain what, and why, different things are noticed and what the associated implications are for learning. Here, *reflection*, as initially characterized by Dewey (e.g., 1933), plays an important role in this process:

> 'Active, persistent and careful consideration of any belief or supposed form of knowledge in the light of the grounds that support it and the further conclusions to which it tends, constitutes reflective thought' (Dewey, 1933, p. 9).

What Dewey saw as reflective thinking involved situations that created (1) 'a state of doubt, hesitation, perplexity, mental difficulty, in which thinking originates', and (2) 'an act of searching, hunting, inquiring, to find material that will resolve the doubt, settle and dispose of the perplexity' (p. 12). The demand 'for the solution of a perplexity, is the steadying and guiding factor in the entire process of reflection.' (p. 14). Reflection is a process, which through the deliberate action of thinking, involves our existing knowledge and viewpoints being changed or informed (Kemmis, 1985):

> 'In reflection we choose, implicitly or explicitly, what to take for granted and what to treat as problematic in the relationships between our thought and action and the social order we inhabit' (p. 148).

Here, we see that both the definition by Dewey (1933) and the ideas of Kemmins (1985) can be said to be congruent with the idea of philosophical contemplation[1]. For the reflective setting Schön draws on Dewey and his idea of transaction to characterize the relation of knower to the known as follows (Schön, 1983):

> 'The inquirer's relation to this situation is transactional. He shapes the situation, but in conversation with it, so that his own models and appreciations are also shaped by the situation. The phenomena that he seeks to understand are partly of his own making; he is in the situation he seeks to understand' (p. 150).

Thus, reflection is personal and different for different persons; an expert makes different reflections on an experience than a student does. All humans reflect and Schön (1983), starting from Dewey's *knowing-in-action*, modelled reflection in two different ways: *reflection-on-action* and *reflection-in-action* (cf. Eraut, 1995). *Reflection-on-action* involves thinking back on the solving of the problem, and *reflection-in-action* involves being aware of, and communicating to others, ones thoughts while engaging in solving the problem. In conclusion, adding *reflection* to *noticing* is to characterize changes in thinking as a function of learning (Mason, 2002). This new thinking becomes the 'seed' needed to construct new meaning from the experience, or re-construct old meaning to see it in new ways (Marton & Booth, 1997).

---

[1] Reflection carries similarities to contemplation – to admire something and think about it, or, the act of considering with attention (theorization). The concept of contemplation is well known from philosophers such as Plato, Aristotle and Plotinus, has been historically applied to the natural sciences (see, for example, Galili, 2013), and has been widely used in religious contexts, often in combination with meditation.

Using this framework and following Eriksson et al. (2014) we define *disciplinary discernment* as noticing something, reflecting on it, and constructing meaning from a disciplinary perspective.

In this article we link disciplinary discernment to the concept of *disciplinary affordances* of representations (Fredlund, Airey, & Linder, 2012). Fredlund et al. (2012) defined *disciplinary affordances* as the 'inherent potential of a representation to provide access to disciplinary knowledge'. Here learning can be framed in terms of discerning the intended meaning of the representations used by the discipline. From this definition we argue that the *disciplinary affordances* of a representation are defined, and given, by the discipline. This implies that when a person observes a representation of, for example, an astronomical object, this representation offers a certain potential to provide access to disciplinary knowledge, that is, it has certain disciplinary affordances. However, the *discerned disciplinary affordances* of a representation are different for different students and constitute a subset of the total disciplinary affordances, set by the discipline community, of that representation (cf. Podolefsky & Finkelstein, 2008).

**Research question**

It has been shown that simulations have great potential for teaching and learning in astronomy because of their ability to make information on the three-dimensional nature of the universe available to students (Eriksson et al., 2014; Joseph, 2011). In order to better understand the role that such simulations play it is interesting to document what lecturers (experts) and university students (novices) discern, when engaging with the same simulation. This led us to our research questions for this paper:

1. What is the discernment reported by university students and lecturers of astronomy when they engage with the same disciplinary representations?

2. How can this discernment be characterized from an educational perspective?

**Method**

The results reported on here come from a larger study, part of which was recently published (Eriksson et al., 2014). The method description is similar for both articles.

*Selection of simulation and the pilot study*

The simulation video for our research project needed to present a realistic journey through our Galaxy and beyond, and include disciplinary-specific representations of objects found in the galaxy. After searching for appropriate simulations in the literature that fulfilled our requirements, the decision fell on the simulation made by the well-known cosmologist Brent Tully (Tully, 2012). We chose to use the first 1.5 minutes of this simulation, cut into seven sections, on average about 15 seconds long. Since the simulation presented so much visual information, the length of these sections was deliberately limited to short clips to restrict the possibility of cognitive overload (Chandler & Sweller, 1991; Mayer & Moreno, 2003). To explore how well these clips worked as a research tool, the sections were piloted on several groups of students and experts before final decisions were made on the cutting of the section clips. This was done by first asking the participants to write down discernment descriptions for each of the sections. Then, their descriptions were analysed using the methodology described below, and finally a survey was produced for on-line use. This was tested and evaluated by disciplinary experts, and minor changes were made before embarking on the main data collection.

*Data collection*

The on-line web survey begins with an ethical agreement and some initial background questions on items such as gender, educational level, and educational setting. It then continues with the seven simulation sections, where the participants are asked the same open-ended questions for each of the seven sections (clips):

1. Please write what comes to mind when you watch this clip, like things you noticed, sudden new realizations or connections, surprising or confusing things.
2. What, if any, "I wonder..." questions did this clip raise for you?
   If you have not noticed something new, feel free to say so.

To reduce test-stress the different simulation sections could be re-played without restrictions (Sieber, 1982). This also facilitated the participants being able to discern as much as they could. After the seven clips were viewed and the two questions for each clip were posted, some follow-up questions were asked. This was done to allow the participants to clarify details that they might have discerned and would like to further address (see Appendix for details).

*Recruitment of participants*

The design of this study called for participants from all university educational levels; from first-year undergraduate students to graduate students and lecturers. In order to reduce effects from local syllabus and educational influences, we recruited participants from a wide variety of international educational contexts. We contacted astronomy lecturers at large universities or national centres for astronomy studies in North America, Europe, Australia and South Africa and asked them to help us obtain access to suitable participants. In total 137 participants from 9 countries (79 men and 58 women) participated in the study. The undergraduate students were divided into two broad educational groups, first-year undergraduate students and post first-year undergraduate students. This was done to accommodate the many variations that we found in the structures of the undergraduate systems for astronomy education across the countries that we obtained our data from. For example, in countries like Australia and South Africa the undergraduate (bachelor) degree consists of three years plus a fourth Honours year (master degree), where astronomy is often given for the first time. In the USA astronomy service courses are first year courses, while astronomy major courses typically only begin in the third year, after two years of mathematics and physics have been successfully completed.

The distribution of educational background of the participants is given in Table 1. The two largest groups of participants were first-year undergraduate students and university lecturers. The first-year undergraduates included students taking introductory courses in astronomy, where no special educational demands for mathematics or science are required. From the survey we also had 9 'Others', these were excluded from the data set since their self-described educational background was difficult to classify. The participants' descriptions were for the most part written in English, however, a Swedish version of the survey was also available for the Swedish participants and their descriptions were translated.

**Table 1.** The number of participants in terms of self-reported location in the higher education system.

| Educational level | Number of participants |
|---|---|
| First-year undergraduate students | 56 |
| Post first-year undergraduate students | 22 |
| Graduate students | 11 |
| Teaching lecturers | 39 |
| Others | 9 |

*Analysis*

A hermeneutic, interpretative approach was chosen, which is widely considered to be appropriate for this type of qualitative educational research (see, for example, Eriksson et al., 2014; Brown, 1996, 2001; Butler, 1998; Case, Marshall, & Linder, 2010; Gallagher, 1991; Seebohm, 2004). This involved using a constant comparative method (Glaser & Strauss, 1967; Strauss, 1987) as follows. The written descriptions from the participants were first read and re-read many times by each of the authors to get a feel for the whole. The qualitative data analysis software NVivo™ was used in the analysis process because of the extensiveness of the data. In the spirit of hermeneutics the leading author started to code a subset of the data and categories began to emerge. Throughout this process, the identified 'parts' were repeatedly compared with the 'whole' data though the process of the hermeneutic circle. The other authors independently checked the emerging categories. This checking included repeated discussions about consistency and accuracy. The rest of the data was then brought into the process, which involved the leading author continuing with the iterative, constant comparative, cyclical approach, consistent with the hermeneutic method. This continued until 'saturation' was reached (Guba & Lincoln, 1985, p. 350), which ended with the formulation of five qualitatively different categories. At this point the other authors took samples of the data and independently sorted it in terms of these five categories. There was extremely good agreement so only very minor changes were made.

**Results**

***Research question 1: What is the discernment reported by university students and lecturers of astronomy when they engage with the same disciplinary representations?***

The analysis for the first research question involved developing a set of qualitatively different categories. This process was broadly thematized in two ways. First in terms of *what was noticed (what-*perspective*)*, and second in terms of *how the participants interpreted* that noticing for meaning (*why-*perspective). The labelling of the categories characterise their central discernment attributes. At this stage we need to point out that the illustrative descriptions that we provide below cannot in themselves capture all the attributes of a given category; what is presented illustrates a selection of the most salient aspects of a given category. We refer to the seas *categories of discernment*. In accordance with the theoretical frame, a given participant's discernment may be characterised by more than one of our categories.

Below we give a short description of the most salient aspects of each of the five category of discernment together with illustrative excerpts from the data set:

*Non-disciplinary Discernment* – In this first category the discernment is restricted to participants noticing different disciplinary representations presented in the simulation, usually without them being able to identify what it is they see. The participants may signal this by posing a question or wondering about *what* it is they notice. Thus, this category serves as the pre-entry level and forms the baseline for any discernment—the participants' attention has been caught by the disciplinary representations shown in the video, for example:

> *I don't know what I see, but it gets brighter and I see horizontal irregularly shaped columns. The horizon is a mixture of dark and bright material, and I have a feeling that there is something bright behind it.*

> *What´s the yellowish band? The horizon-looking thing. And what´s the cloud-looking things in it?*

> *Journey's getting farther into space - objects getting closer; sudden appearance of a distinct narrow opaque object in the smaller moving structure; what could this be?*

> *What are the red clumps?*

As these excerpts illustrate, the participants notice different structures but do not know what it is that they are 'seeing'. This is the first step in the process of learning; the experience offered by the simulation has caught their attention and they express the noticing of different representations of 'things', and begin to reflect on what these might be.

*Disciplinary Identification* – The disciplinary discernment at this level involves naming, or recognising, the most salient disciplinary objects in the representation.

This category represents the first signs of disciplinary discernment, as we define it, related to astronomical phenomena and recognition of astronomical structures. In other words, focusing on parts and distinguishing what these afford from a disciplinary perspective. In this category the participants are thus identifying what it is they notice. Almost all the participants give data that can be coded in this category. In this, we see that many descriptions move from '-What is that?' into '-Oh, that is …', revealing reflective awareness on sameness and differences (Marton & Booth, 1997) of the structural components of the Universe and how these are represented in the simulation clips. Four examples of such descriptions are:

> *We see a beautiful spiral galaxy! I saw the bulge and the spiral arms, and as it zoomed out I could see globular clusters around the galaxy.*
>
> *Surprised and pleased that I was correct with my guess that it was a spiral arm we were looking at. Though I could not reconcile it with what appeared to be the nucleus of the galaxy. This movie made it all clear. We were panning across the spiral arm of a galaxy, which then led to the centre of the galaxy. It is a very nice shot.*
>
> *I'm travelling through the Milky Way Galaxy, towards the stars that makes up the constellation Orion.*
>
> *Neat picture of a galaxy with the dark dust obscuring the light from behind and the lighter gas clouds on the edges. I wonder how far that spiral arm is beyond the Orion Neb, and where the centre of the galaxy lies with respect to the arm.*

In this category the first signs of recognition of disciplinary-specific representations emerge. The participants express discernment of what the different structures are and what they represent. Thus, the descriptions in this category reveal identification of the representations of astronomical objects at different scales.

*Disciplinary Explanation* – The disciplinary discernment for this level involves explaining or assigning disciplinary meaning to the discerned disciplinary objects, i.e. 'discover' the affordances of the representations.

In this category we see a shift in the description from the *what*-perspective towards a *why*-perspective. We recognise this as a major step in the participants' disciplinary discernment; they start to use their disciplinary knowledge to try to interpret what they see in terms of astronomical properties and astrophysical processes. For example, it could be discernment related to composition (structural aspects or what the different objects are made of), colour (in relation to emission, absorption, and/or temperature), or other astrophysical aspects (including processes). Four examples of such descriptions are:

> *My only guess […] would be to say that these are nebula. They are probably made out of helium or hydrogen gasses.*
>
> *… appears to be red from strong Balmer lines, but I am under the impression that the Orion nebula appears slightly green to the naked eye due to trace amounts of ionised oxygen.*
>
> *I wonder how big the star will actually become due to gravity and pressure.*
>
> *Interesting to realize how gravity holds the galaxy together.*

As these excerpts highlight, participants' disciplinary knowledge is used to try to explain what is discerned to construct an understanding of why things appear as they do in the simulation. We thus see that the disciplinary affordances of representations are beginning to be 'discovered' by the participants.

*Disciplinary Appreciation* – The disciplinary discernment making up this category involves analysing and acknowledging the value of the disciplinary affordances of the representation.

This category represents a more advanced level of disciplinary discernment in terms of it bringing together all previous categories to generate a more holistic view of the galaxy, including its different representations of stellar objects and how they work together at different levels of detail. This calls for the ability to discern and analyse the disciplinary affordances of the representations at all levels. Such ability made it possible for the participants to appreciate the simulation in different ways. Below follow four examples of descriptions:

> *Looks like the compact source seen earlier has resolved into an HII region with star cluster. Looks like the Rosetta Nebula. So pretty. Interesting how the stars seem to have carved themselves out a hole in the middle.*
>
> *Out of the plane of the galaxy, which we can now see as a spiral galaxy much like the Milky Way. There are a number of pink regions that look like they are forming new stars, and these regions seem to lie along radial spiral streaks with (older?) yellow regions between them.*
>
> *So we were between spiral arms. It seems crowded - lots of stars and gas. It is hard to appreciate the stellar neighborhood when we have talk about the distances to our nearest star at 4.2 light years. That seems very far away, yet looking at this rich neighborhood, on the stellar scale, it is actually very close.*
>
> *When I see that clip I start to think about all the things I have learned during the course. What a nebula is, how stars are born, supernovae, and other concepts that I have learned. This picture is not entirely like other pictures I have seen on this object.*

These excerpts highlight ways in which the participants 'appreciate' the representations, by combining disciplinary knowledge from different areas within the discourse of astronomy, to build a holistic understanding of what the representations are intended to afford.

*Disciplinary Evaluation* – This category characterises the most advanced level of disciplinary discernment found. The discernment involves analysing and critiquing the representation used for an intended affordance. The critique could involve both positive and negative elements as seen in the five descriptions below:

> *It looked like there was a cluster of stars at the bottom of the screen near the end. Before I saw that I thought this video was mis-representing* [sic.] *the contents of a galaxy by only showing nebula.*
>
> *The star forming regions also became more evident as the view moved closer to Orion. One nice thing about the movie is that the star forming regions did not get noticeably brighter as we moved closer, which makes sense since surface brightness should be conserved.*
>
> *I was disappointed that it took roughly the same amount of time to pass through it as it took to pass through the Orion Nebula--missing a teachable moment on showing a major difference between the two.*
>
> *I wonder if all the relative star positions are correct, if the scale is correct. Probably not as the computation required would be excessive. I wonder what percentage of the data is real. I'm also wondering if this simulation takes into account blue shift, since the camera is moving toward the stars.*

> *The clip implies that the Orion nebula is a discrete thing, and that one can "see" the Milky Way around its edge. But the nebula is actually only a small part of a much larger opaque cloud – so this part is a bit misleading. There is a lot going on that we can't see at visible wavelengths – this would be a very different clip in infrared light, for example.*

The descriptions in this category reflect high levels of disciplinary knowledge and include certain degrees of critique or criticism. This is different from the precious category, where only appreciation could be seen in the descriptions. Furthermore, some of the descriptions in this category also include aspects related to using such a resource in the teaching practice of the discipline.

These five categories represent the different ways in which disciplinary discernment can be characterised as the answer to the first research question. It became very clear when analysing the data that the simulation clips offer very different discernment possibilities for the different participants. Under the reasonable assumption that higher educational levels imply more and deeper disciplinary knowledge, the analysis can be seen to show that the more disciplinary knowledge the participants have, the less non-disciplinary discernment they reported. Figure 1 illustrates this by showing how less attention was paid to non-disciplinary characteristics as the educational level increased.

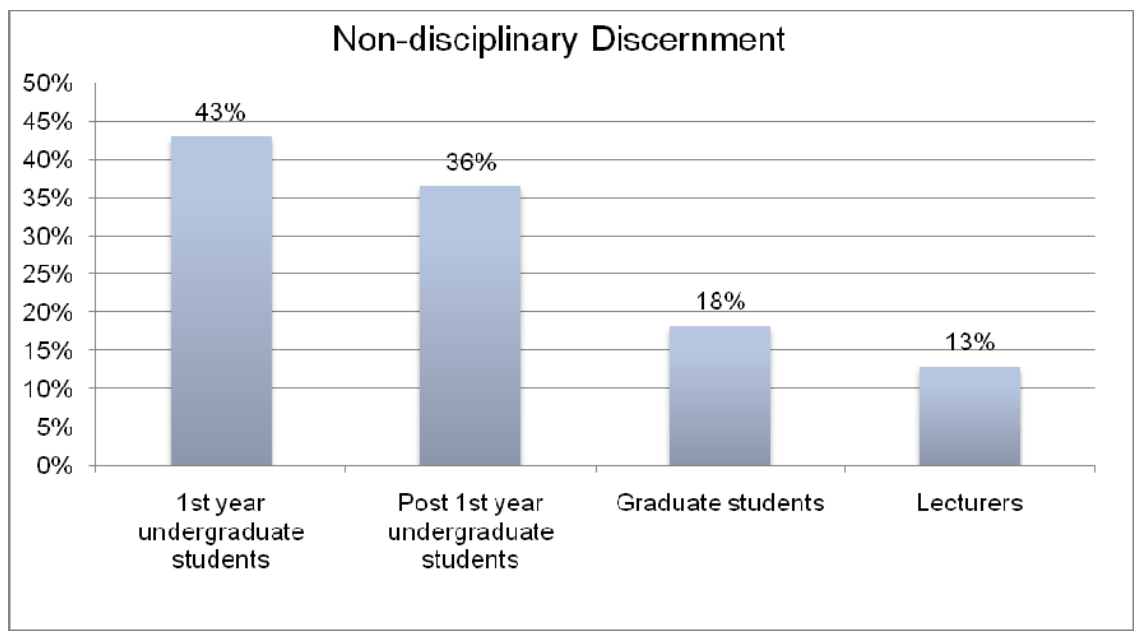

Figure 1. The percentages of participants who provided descriptions of discernment, illustrating how less attention gets paid to non-disciplinary characteristics as the education level increases.

### *Research question 2: How can this discernment be characterized from an educational perspective?*

As a consequence of the results for the first research question, we now use the identified and described categories, that characterise the participants' reported discernment, to construct what we call an *Anatomy of Disciplinary Discernment* (ADD), see Table 2. The ADD encapsulates the increasing complexity of intended meaning of representations, what we refer to as disciplinary affordances, and the categories can be seen as a hierarchy of discernment progress. It describes *the ways in which the disciplinary affordances of a given representation may be discerned*. This discernment involves accessing disciplinary knowledge to assign meaning to a representation. Therefore, *disciplinary knowledge* can be said to be a decisive factor for the possible discernment, hence meaning-making, offered by the representation. The unit of analysis for our ADD is thus *the discernment of disciplinary affordances of the representations being used in the simulation*.

Table 2: The *Anatomy of Disciplinary Discernment*. For details and description of the categories, see the text.

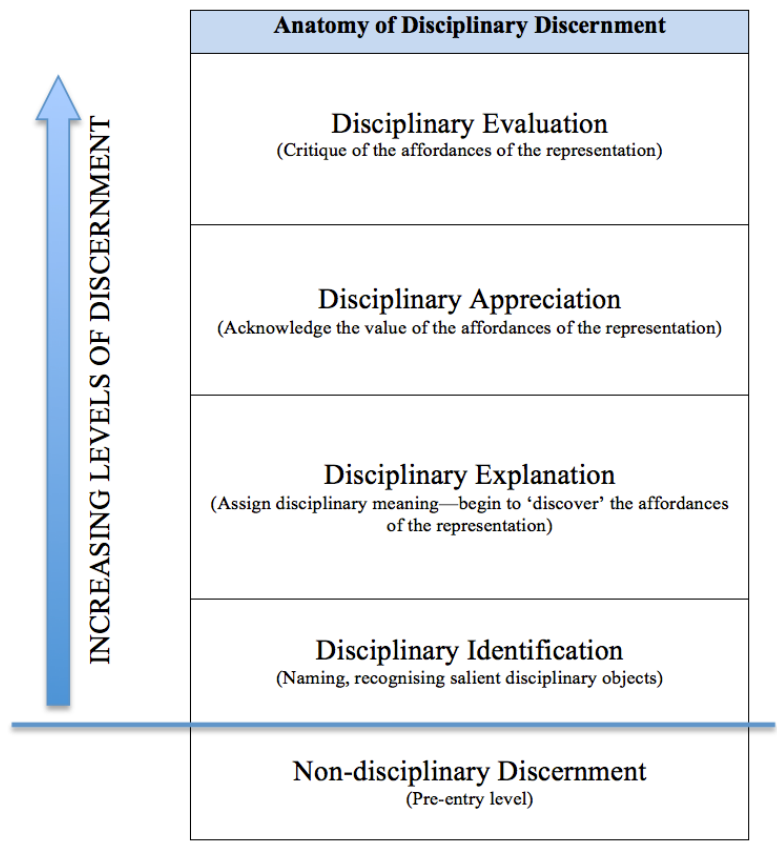

These categories can be seen to form a hierarchy of discernment that begins after a baseline of simple noticing. This baseline is characterized as non-disciplinary discernment. An illustrative example is: suppose new-to-astronomy students are shown images of a nebula in their introductory class meeting. The students are sure to notice a diversity of colour but that colouring is unlikely to convey any astronomical significance to them beyond looking nice. The first level of the ADD (Disciplinary Identification) is characterized by the recognition and naming of salient objects. An illustrative example is: being able to identify the structured coloured regions in an astronomy image as representing different nebulae and using that discernment to recognise a particular nebulae structure sufficiently well enough to name it. In the second level of the ADD (Disciplinary Explanation) disciplinary meaning is assigned to a representation. This means that a "discovery"— a new awareness — of the disciplinary affordances of representations takes place. An illustrative example is: getting to see the coloured details in terms of interstellar cloud of dust, hydrogen, helium and other ionized gases that can manifest as star-forming regions. It is a step beyond the previous category in that disciplinary knowledge is being used in new ways to make these kinds of connections. The third level of the ADD (Disciplinary Appreciation), involves coming to see the value of the representation for the discipline—appreciating its disciplinary "power". Example: connecting the different types of colour to interstellar cloud of dust, hydrogen, helium and other ionized gases in ways that account for stellar birth and evolution. It brings together different parts of disciplinary knowledge, for instance radiation, gravitation, gas laws, thermodynamics, nuclear physics, etc. In other words, a growing appreciation of how the parts of a nebula can work together to generate a snap-shot of stellar evolution. Disciplinary Appreciation calls for accessing disciplinary knowledge at many different levels of detail simultaneously in order to appreciate a disciplinary representational image of, say, a nebula. It is here that a value of the disciplinary affordances becomes part of the discernment. Finally, the highest level of the ADD (Disciplinary Evaluation) brings being able to critique the representations into the discernment. Such disciplinary discernment is the hallmark of disciplinary expertise. An illustrative example is: a discernment that includes being able to see a particular representation of, say, a nebula being limited in terms of the representations being used. For example, representations of nebulae using colours cannot on their own provide full and holistic disciplinary affordances.

However, we want to take the ADD even further to be able to answer the second research question on how this discernment can be characterized from an educational perspective. To do so, we need to frame the ADD around the concept of learning by involving two powerful educational ideas; the 'Spiral Curriculum' (Bruner, 1960) and 'Visible Learning' (Hattie, 2009; 2012).

Bruner's (1960) notion of the spiral curriculum involves information being structured so that complex ideas can be taught at a simplified level first, and then re-visited at more complex levels later on. Therefore, subjects would be taught at levels of gradually increasing disciplinary representation affordance (hence the spiral analogy). Ideally, teaching this way should lead to students being able to participate in all of the ADD categories. However, this spiral curriculum idea does not explicitly address how this can occur. We interpret our results as a growth into the discipline that could be described in similar ways as to Bruner's spiral curriculum approach, see Figure 2. This representation of our result shows that, ideally, for each turn in the spiral, the student's disciplinary discernment would *cross a category boundary* and move to the next level of the ADD, as the disciplinary knowledge increases through the process of learning. Bruner's work also suggests that all students are capable of learning any material so long as the teaching is organized appropriately, the ADD provides the framework for doing this. Bruner (1961) proposes that students' construct knowledge for themselves. That is, they cannot be "given" it. Such construction takes place by a process involving organising and categorizing information. Our proposal is that this organising and categorising should take place through being provided with educational experience that follows sequencing based on the ADD. That is, students are to be given the opportunity to "discover" the disciplinary way of organising and categorizing things rather than just being given by teachers. This concept of discovery learning implies that students construct their own knowledge for themselves (also known as a constructivist approach). However, this may not be sufficient in itself and guidance is often required (see, for example, Mayer, 2004). Here, Hattie's (2009; 2012) idea of visible learning becomes important and highlights the roll of the teacher in this process: 'Visible Learning and Teaching occurs when teachers see learning through the eyes of students and help them become their own teachers' (Visible learning, 2014). We will address this in more detail in the Conclusion and Implications section below.

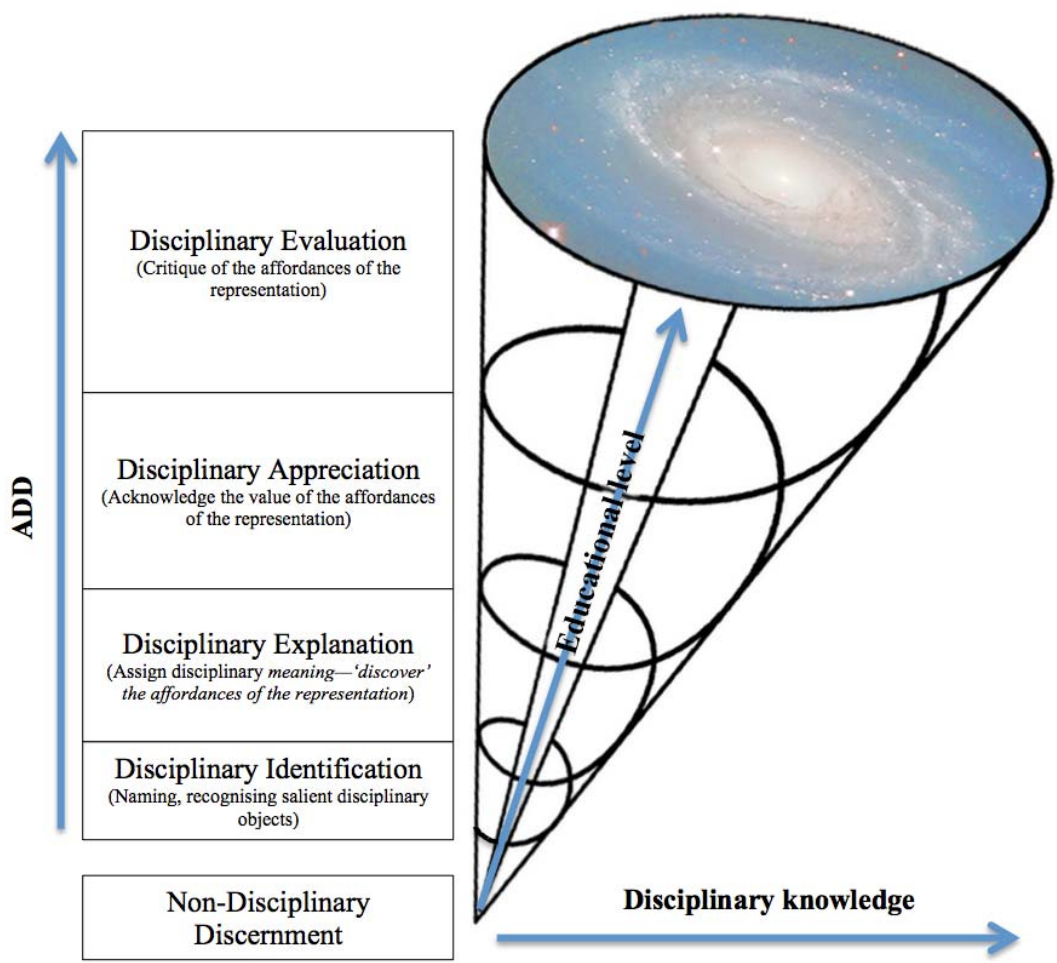

Figure 2. An idealized visual representation using our proposed ADD combined with Bruner's spiral curriculum to show the teaching and learning trajectory of a science student. Through the iterative revisiting of different material, disciplinary knowledge increases (illustrated by the width of the spiral) together with the ability to discern disciplinary affordances of representations. This reflects the movement upward through the proposed levels of the ADD. Note that each turn of the spiral is used to illustrate the crossing of a category boundary to a new level of the ADD.

**Discussion**

What we have found is a framework describing *what* and *how* different disciplinary representations are discerned in a disciplinary manner. The proposed ADD describes the developmental characteristics of the ability to discern disciplinary affordances of representations in a way that has not been done before in the literature. This framework could be used to explain what Sherlock Holmes could so easily do whereas Dr Watson could not; *the difference in competence lies in making the relevant disciplinary discernment*, described by the different levels of the ADD. Disciplinary experts, like Sherlock Holmes, have developed competences in applying different strategies to interpret discerned details from different representations (Ertmer & Newby, 1996). It is therefore not surprising to find that astronomy graduate students and astronomy lecturers generally are good at discerning the relevant disciplinary-specific representations in the simulation. They have developed competences similar to what Goodwin (1994) calls *professional vision*. This enables them to discern many more aspects due to their educational background and disciplinary knowledge, than the undergraduate students could possibly do at their present level of astronomy education and disciplinary knowledge. As such, we expect to find that many of the graduate students and lecturers discern aspects of the simulations that can be referred to the highest levels of the ADD. We find that they have developed

'sensitivity to patterns of meaningful information that are not available to novices' (Bransford, Brown, & Cocking, 2000, p. 33) in that they can; *evaluate* the simulation by observations and discernment of relevant details, even if these are not directly visible, make connections and *criticise* the simulation in a relatively effortless, or automatic, manner (Eberbach & Crowley, 2009; Schneider & Shiffrin, 1977). The undergraduate students, on the other hand, are often found to focus on the 'wrong things' in the representations, as clearly shown in Figure 1.By learning to discern, an 'outsider' eventually becomes an 'insider', or expert (cf. Podolefsky & Finkelstein, 2008). This learning process leads to disciplinary knowledge, gathered through experiences and variation, and hence disciplinary knowledge must be a decisive factor for disciplinary discernment. Through the process of acquiring more disciplinary knowledge, the better the disciplinary discernment will be and hence the higher in the ADD a student is likely to be found. Our results suggest that *disciplinary discernment is intertwined with disciplinary knowledge*, as illustrated in Figure 2, supported by the finding that less attention gets paid to non-disciplinary characteristics as the education level increases, see Figure 1. However, it is in the application of the ADD that its 'power' becomes evident.

**Limitations**

Not being a longitudinal study could be considered to be a limitation of our work. The data consists of snap-shots taken at a series of broad educational levels, which arguably provides a data set that is more appropriate for a competency framing since these snap shots provide an instantaneous cross-level picture. A second possible limitation is the use of a single simulation. However, this simulation includes a large variety of representations of objects in the Universe, which brings credence to its validity for the study. Planned, future research will explore the validity of the ADD using other types of representations, for example graphs and diagrams. Finally, the large number of participants (N=137), in combination with the chosen methodology, provides a validity for the categorisation, and hence the ADD.

**Conclusions and implications**

This study aimed at finding and describing, from a disciplinary perspective, what students and lecturers discerned when engaging with a simulation video. This led to us developing the ADD; which, from an educational perspective is how disciplinary knowledge can be seen to increase as a function of a growing ability to discern disciplinary crucial aspects from a vast array of potential affordances of a given representation. The ADD hierarchy thus mirrors the development from disciplinary 'outsider' to disciplinary 'insider', similar to Bruner's spiral curriculum idea (see Figure 2). However, it may not be immediately clear from the combined ADD and spiral curriculum resemblance the ways in which the described hierarchy could be used. It is here that Hattie's idea of *visible learning* becomes central: the role of the teacher is crucial for the success of the students in crossing category boundaries in the ADD: 'It is teachers seeing learning through the eyes of students [and,] the greatest effects on student learning occurs when teachers become learners of their own teaching, and when students become their own teachers' (Hattie, 2012, p.14). We find this particularly important when it comes to discernment of disciplinary affordances of representations: teachers need to find out where their students are in the ADD and then teach them to discern relevant disciplinary affordances of representations (cf. Ausubel, Novak, & Hanesian, 1978). This implies that lecturers should begin teaching sequences with activities which draw out students' ideas and provide opportunities for expression of those ideas (Rowell, 1998).

We believe that our model provides a pragmatic guide for how to approach teaching and learning in effective ways using the ADD. In the following we give a short illustrative example of how such a teaching sequence could be organised. Assume a teaching sequence concerning galaxy rotation and rotation curves. A lecturer would first need to find or construct a simulation that clearly and realistically represents a rotating spiral galaxy and critically evaluate this simulation for its disciplinary affordances (Evaluation level). Here, the lecturer must become a learner of her own practice and try to envision what her students will discern from the simulation. After deciding on the usefulness of the simulation, it is presented for the students. These are asked to study it carefully and answer to questions on what they notice. We suggest that the students express their answers in written. The lecturer can now organise the answers using the ADD to find out at what disciplinary discernment level the students are. Now the lecturer can plan the teaching sequence using this new knowledge and start at the students' level. Assume that the students are found poor at discerning disciplinary affordances connected to galaxy rotation (Non-disciplinary discernment level). Then, for example, start by discussing the experience of rotation itself by pointing it out for the students (Identification level). Then go back to the simulation and help the students to discern how the galaxy is rotating at different radii. From this discernment it may become relevant for the students to explore how this rotation can be expressed and explained, using observations and laws of physics (disciplinary knowledge) to construct a rotation curve, i.e. they become their own teachers. This expanded disciplinary knowledge may enable the students to discover more disciplinary affordances by the simulation when revisiting it again, for example the role of dark matter, and hence cross the boundary into the Explanation level. From applying their new disciplinary knowledge to more galaxies, the students may be able to appreciate the represented galaxy's disciplinary affordances and start to construct advanced disciplinary knowledge, including the concept of dark matter, on rotation of spiral galaxies (Appreciation level).

In conclusion, the proposed ADD is a framework for how to see teaching and learning science using disciplinary-specific representations. The first aspects to consider for a lecturer would then be to realise that the students are not likely to discern the same things as the lecturer does from a representation. What is 'obvious' to disciplinary experts may not even be discernable to those still outside or 'moving into' the discipline (Eriksson et al., 2014; Northedge, 2002; Rapp, 2005; Tobias, 1986). In fact, it is often found that disciplinary experts have lost the ability to see things as students might see them (Bransford et al., 2000). The second aspect to consider would be in the application of the ADD; lecturers should focus on activities that are designed to help students to *cross over the category boundaries* in the ADD. Northedge highlights the importance of this process: 'the teacher guides the students on an excursion into the target discourse arena, gradually shifting the frame of reference until it corresponds well enough to allow sense to be made within the specialist discourse' (Northedge, 2002, p. 263), signalling the mapping between disciplinary-specific representations without over-burdening the students by making this task too complicated (Ainsworth, 2008). We suggest that it is only then that students can be expected to discern and discuss details of the representations used in disciplinary discourse, i.e. take control of their own learning and become their own teachers. We argue that by referring to the ADD, a lecturer who has determined what students discern and who understands the disciplinary affordances of the material at hand is better empowered to devise appropriate learning experiences. Thus, it is the modelling of the role of the lecturer as one of facilitating boundary crossing in the Anatomy of Disciplinary Discernment that we believe is the major contribution of the work presented here.

**Acknowledgement**
First, we would like to thank Brent Tully for granting permission to use his highly acclaimed video "Flight to the Virgo Cluster" in our research project. Next, thanks to Ed Prather for help with the distributing and publicizing the survey in North America. Also, many thanks to Anne Linder, who critically reviewed the article for language issues and consistency and suggested valuable improvements. Finally, we would like to thank everyone—students and faculty—who took the time to watch the video and answer the survey.

Wallace, C. S., Prather, E. E., & Duncan, D. K. (2012). A Study of General Education Astronomy Students' Understandings of Cosmology. Part IV. Common Difficulties Students Experience with Cosmology. *Astronomy Education Review, 11*(1), 010104-010111.

**Appendix**

Below are the questions found in the survey:
1) *Please, write your name.*
2) *Gender?*
   - *a. Male*
   - *b. Female*
3) *What is your first language?*
   - *a. Swedish*
   - *b. English*
   - *c. None of these*
4) *At what university do you study or work?   (city and country)*
5) *What is your main educational background?*
   - *a. First year astronomy undergraduate student*
   - *b. Second (or more) year astronomy undergraduate student*
   - *c. Astronomy graduate student*
   - *d. First year physics undergraduate student*
   - *e. Second (or more) year physics undergraduate student*
   - *f. Physics graduate student*
   - *g. Introductory level astronomy student*
   - *h. University teacher in astronomy*
   - *i. University teacher in physics*
   - *j. School teacher*
   - *k. Other*
6) *Watch this clip and answer the questions in the provided box below! Please answer the questions in order of appearance, using numbers. 1)*
   - *a. Please write what comes to mind when you watch this clip, like things you noticed, sudden new realizations or connections, surprising or confusing things.*
   - *b. What, if any, "I wonder..." questions did this clip raise for you?*
     *If you have not noticed something new, feel free to say so. When finished with all questions, click the next button.*

For the first clip, we also asked if it was clear where the journey started.

After the seven clips, follow-up questions were asked:
14) *Now that you have seen the whole movie, did you get a good sense of where the journey started and where it ended? Please explain as fully as possible.*
15) *With respect to the movie, mention those things that particularly caught your attention and explain why (e.g. new things that you noticed, things you noticed differently now when you have seen the whole movie, or things that you found amazing to notice).*
16) *What, if any, new connections between phenomena or structures did the movie as a whole make for you? Please explain as fully as possible.*
17) *What, if anything, surprised you in the movie as a whole?*
18) *How realistic did you feel that the movie was? Please explain your reasoning.*